\documentclass[a4paper,11t]{article}

\usepackage[top=2.54cm,bottom=2.54cm, left=2.54cm, right=2.54cm]{geometry}

\usepackage{amsmath}
\usepackage{amsfonts}
\usepackage{amssymb}
\usepackage{graphicx}
\usepackage{bm}
\usepackage{enumerate}
\usepackage{color}
\usepackage[sort,comma]{natbib}
\usepackage{float}
\usepackage[colorlinks=true, allcolors=blue]{hyperref}
\usepackage{multirow}
\usepackage{booktabs}

\title{Computationally efficient univariate filtering for massive data}
\author{Tsagris M., Alenazi A. and Fafalios S. \\
\\
University of Crete, Department of Economics, Gallos Campus, Rethymnon, Greece \\
\href{mailto:mtsagris@uoc.gr}{mtsagris@uoc.gr}  \\
Department of Mathematics, Northern Border University, Arar, Saudi Arabia, \\ \href{mailto:a.alenazi@nbu.edu.sa}{a.alenazi@nbu.edu.sa} and \\
University of Crete, Department of Computer Science, Voutes Campus, Herakleion, Greece \\
\href{mailto:stefanosfafalios@gmail.com}{stefanosfafalios@gmail.com} 
}

\begin{document}

\maketitle

\begin{abstract}
The vast availability of large scale, massive and big data has increased the computational cost of data analysis. One such case is the computational cost of the univariate filtering which typically involves fitting many univariate regression models and is essential for numerous variable selection algorithms to reduce the number of predictor variables. The paper manifests how to dramatically reduce that computational cost by employing the score test or the simple Pearson correlation (or the $t$-test for binary responses). Extensive Monte Carlo simulation studies will demonstrate their advantages and disadvantages compared to the likelihood ratio test and examples with real data will illustrate the performance of the score test and the log-likelihood ratio test under realistic scenarios. Depending on the regression model used, the score test is $30 - 6,000$ times faster than the log-likelihood ratio test and produces nearly the same results. Hence this paper strongly recommends to substitute the log-likelihood ratio test with the score test when coping with large scale data, massive data, big data, or even with data whose sample size is in the order of a few tens of thousands or higher. 

\paragraph{keywords:} Univariate filtering, univariate regression models, computational efficiency
\end{abstract}

\section{Introduction}
Massive data, which require high computing power, have become a frequent phenomenon nowadays. Reducing the computational cost entailed by massive data, using computationally efficient algorithms, is beneficiary for research and industry related purposes. In bionformatics for instance, analysis of numerous gene expression data that contain 55,000 variables and in computer science, analysis of big data (order of Terabytes and higher) are common tasks. Computationally efficient algorithms are also highly desirable and required by banks, large scale institutions and companies that handle big data because those algorithms not only reduce the waiting time but further have an economic impact since they can reduce electricity expenses. 

A common task met in both research and industry is variable selection (VS), described as follows. When a response variable $Y$ (for example a phenotype, disease status, survival time) is given along with a set $\bf X$ of $d$ predictor (or independent) variables, both consisting of $n$ observations, VS attempts to identify the minimal set of predictor variables whose predictive capability on the response is optimal. In bioinformatics for instance, the goal is to identify the genes whose expression levels allow for early diagnosis of some disease \cite{tsamardinos2003towards}.

Over the years, there has been an accumulation of VS algorithms in many data science fields, such as bioinformatics, statistics, machine learning, and signal processing. Most algorithms tackle the VS problem from an agglomerative, forward selection perspective. They commence with an empty set of variables and move forward by adding one or more variables at each time. Max-Min Parents and Children \citep{tsamardinos2003b}, Statistically Equivalent Signatures \citep{lagani2017}, Forward Backward with Early Dropping \citep{borboudakis2019}, Orthogonal Matching Pursuit \citep{chen1989,pati1993,davis1994}, Sure Independence Screening \citep{fan2008,fan2010}, forward selection \citep{weisberg1980} and forward stepwise regression \citep{weisberg1980} are some examples of VS algorithms that initially perform univariate filtering. At that step the most statistically significant variable, or the variable mostly correlated with the response\footnote{In all cases and examples considered in this paper, only continuous predictor variables will be used.} is detected, while significant variables or the $k$ ($=\frac{n}{\log n}$ for example) most significant variables are retained for further analysis. 

Univariate filtering with continuous responses is fast enough because of the fast implementation of the correlation between $\bf y$ and each of the ${\bf X}_is$. With non-continuous responses though (count data, nominal, ordinal, survival), $d$ univariate regressions and hence $d$ log-likelihood ratio tests must be performed. This can be computationally really heavy with tens of thousands of variables or even with large sample sizes (hundreds of thousands).

Statistical softwares, such as R, are not computationally efficient in fitting numerous regression models when built-in commands are applied, such as \textit{glm} or any regression model offered by a package, inside a \textit{for} loop. Self implementation of the regression models and employment of parallel computing can assist reduce the execution time in R. The same recipe can be applied with C++, resulting in higher savings\footnote{Numerous C++ regressions models can be found in the R package \textbf{Rfast} (Papadakis et al., 2019).}. This then raises the question of whether \textit{univariate filtering can become more efficient or extremely efficient, and effectively reduce the computational cost of numerous VS algorithms}. The answer is \textbf{Yes: employment of the score test or of the Pearson correlation coefficient allows for computationally extremely efficient univariate filtering}.
Also, specifically for logistic regression, the Welch's $t$-test \citep{welch1951} is another possibility. 

The score test, also known as Rao's test \citep{rao1948} or Lagrange Multiplier test \citep{greene2003}, is robust in the sense that it does not depend on the functional relationship between the response and the predictor variable(s) and it depends on the null distribution of the response $y$ only through the MLE of the distribution under the $H_0$ \citep{chen1983}. It is asymptotically equivalent to the log-likelihood ratio test \citep{greene2003} and for logistic and Poisson regression its formula is similar to the Pearson correlation coefficient \citep{hosmer2013}. Both the score test and Pearson correlation coefficient are applicable to numerous regression models, such as logistic, Poisson, negative binomial, Beta, Gamma, etc. The score test is advantageous because it is computationally cheap. \textit{For logistic regression the score test is between 30 to 70 times faster than a C++ implementation of the log-likelihood ratio test, while with Beta regression it is more than $6,000$ times faster than the log-likelihood ratio test using a Beta regression implementation in R}. Score test's computational efficiency springs from the fact that it fits a single regression model only, under the null hypothesis, unlike the log-likelihood ratio test that requires fitting many regression models under the alternative hypothesis as well. 

The strong point of the score test is its asymptotic correctness (i.e. it requires the sample sizes to be at the order of thousands). The larger the sample size is the more accurate the approximation to the log-likelihood ratio test is and the higher the score test's computational efficiency is. The asymptotic proximity of the two tests can be explained by the fact, that the log-likelihood ratio test and the score test differ by $O_p\left(n^{-1/2}\right)$ \citet{young2005}, where the $O_p()$ notation indicates a random variable that is asymptotically bounded in probability. In addition, both scores are parametrisation invariant\footnote{Parametrisation invariance requires that the conclusions of a statistical analysis be unchanged for any reasonably smooth one-to-one function of $\theta$ \citep{young2005}.}. For a comparison of the log-likelihood ratio test and score test in terms of the expected length of their confidence intervals the reader is referred to \citet{mukerjee2001}. The practical advantages and disadvantages of the score test and of the Pearson correlation coefficient will be  illustrated and conclusions will be drawn, via simulation studies and experiments with real data, at which different types of regression models will be considered. 

The next section presents the log-likelihood ratio test that relies upon fitting regression models, the score test and the Pearson correlation coefficient, followed by reference to related work. Section \ref{simulations} illustrates the computational benefit of the score test and of Pearson correlation against various regression models, including inter comparisons among them in terms of type I error, correlation of the p-values and percentage of agreement of rejection of the $H_0$. Section \ref{real_examples} illustrate the log-likelihood ratio and the score test using real data and finally Section \ref{conclusions} concludes the paper.

\section{Log-likelihood ratio and Score tests for regression models and Pearson correlation coefficient}
Assume a response variable, a $n\times 1$ vector of observations $\bf y$ and a set of predictor variables, an $n \times d$ matrix {$\bf X$}, where $n$ denotes the sample size and $d$ denotes the number of variables are given. At first a regression model with only the intercept is fitted and its log-likelihood is computed. Then for each variable a regression model is fitted: 
\begin{eqnarray*}
& & H_0: g(y) = a_j. \\
& & H_1: g(y) = a_j+b_jx_j \ \ (j = 1, \ldots, d ).
\end{eqnarray*}
Univariate filtering identifies the statistically significant predictor variables, or the $b_js$ that are statistically significantly different from zero.

\begin{itemize}
\item For each regression model in $H_1$ its associated log-likelihood is computed and hence the log-likelihood ratio test statistic is computed by 
\begin{eqnarray} \label{llr}
\Lambda = 2\left( \ell_1-\ell_0\right),
\end{eqnarray}
where $\ell_1$ and $\ell_0$ are the log-likelihood values under the $H_1$ and $H_0$ respectively. Under the $H_0$, $\Lambda \sim \chi^2_1$ \citep{young2005}. 

\item The score function is the derivative of the log-likelihood $U_0 = U\left(\theta_0\right)=\sum_{i=1}^n\frac{\partial \ell\left(\theta_0\right)}{\partial \theta}$, where $\theta$ denotes the value of the parameter of interest and $\theta_0$ denotes its value under the null hypothesis. From standard likelihood theory it is known that $E_{\theta}\left(U_{\theta} \right)=0$ and $Var_{\theta}\left(U_{\theta}\right)=I\left(\theta\right)$, where $I\left(\theta\right)$ is the Fisher information. By an application of the central limit theorem combined with Slutsky's lemma, under the $H_0$, $n^{-1/2}U\left(\theta_0\right)\xrightarrow{d}N\left(0, I\left(\theta_0\right)\right)$ \citep{young2005}, and hence the score test
\begin{eqnarray} \label{score}
S^2 = \frac{U_0^2}{Var(U_0)} \sim \chi^2_1.
\end{eqnarray} 

\item The sample Pearson correlation coefficient is computed by
\begin{eqnarray} \label{r}
r = \frac{n\sum_{i=1}^nx_iy_i - \sum_{i=1}^nx_i\sum_{i=1}^ny_i}{\sqrt{n\sum_{i=1}^nx_i^2 - \left(\sum_{i=1}^nx_i\right)^2}
\sqrt{n\sum_{i=1}^ny_i^2 - \left(\sum_{i=1}^ny_i\right)^2}}.
\end{eqnarray}
Under the $H_0$ (the two variables $X$ and $Y$ are linearly independent), the test statistic $Z = 0.5 \log{\frac{1+r}{1-r}}\sqrt{n-3}$ asymptotically follows a $N(0, 1)$, while for small $n$, $N(0, 1)$ can be substituted by $t_{n-3}$.
\end{itemize}

\subsection{Large sample asymptotics of the score test}
Below is a short proof of the asymptotic equivalence of the score test and log-likelihood ratio test when $\theta$ is scalar, as in the case this paper examines. By expanding the score function $U\left(\hat{\theta}\right)$ using Taylor series about $\theta$ one can obtain \citep{brazzale2007} $$ S\left(\theta\right) = \left(\hat{\theta} - \theta\right)I\left(\theta\right)^{1/2}= I\left(\theta\right)^{-1/2}U\left(\hat{\theta}\right)\left[1+o_p(1) \right].$$
A similar expression for the log-likelihood ratio test gives \citep{brazzale2007}
$$\Lambda(\theta) = \left(\hat{\theta} - \theta\right)^2I\left(\theta\right)\left[1+o_p(1) \right],$$
where $o_p(1)$ indicates a random variable that converges in probability to $0$. This proves that the two tests are asymptotically equivalent \citep{young2005}. 

\subsection{Score test formula for selected regression models}
Formulas of the score test for some common regression models are given below.
\begin{itemize}
\item With binary responses $(0 \ \text{or} \  1)$, logistic regression is usually employed. The log-likelihood of the logistic regression is given by $$\ell_1 = \sum_{i=1}^n\left[y_i\log{p_i}+\left(1-y_i\right)\log{\left(1-p_i\right)}\right],$$ where $p_i=\frac{1}{1+e^{-a-bx_i}}$. The score test takes the following form (Hosmer et al., 2013) 
\begin{eqnarray} \label{logistic}
S_{Bin} &=& \frac{\sum_{i=1}^ny_ix_i-\hat{p}\sum_{i=1}^nx_i}{\sqrt{\left[\sum_{i=1}^nx_i^2-\left(\sum_{i=1}^nx_i\right)^2/n\right]\left[\hat{p}\left(1-\hat{p}\right)\right]}},
\end{eqnarray}
where $\hat{p}=\sum_{i=1}^n\frac{y_i}{n}$. The formula in (\ref{logistic}) is equivalent to the square of the Cochran-Armitage test statistic for testing trends in a single $2 \times J$ contingency table \citep{chen1983}. 

\item With count data, the Poisson regression is the simplest model employed whose log-likelihood is given by $$\ell_1=\sum_{i=1}^n\left[y_i\log{\left(\lambda_i\right)}-\lambda_i-\log{\left(y_i!\right)} \right],$$ where $\lambda_i=e^{a+bx_i}$. The form of the score test in this case is
\begin{eqnarray} \label{poisson}
S_{Pois} &=& \frac{\sum_{i=1}^ny_ix_i-\bar{y}\sum_{i=1}^nx_i}{\sqrt{\left[\sum_{i=1}^nx_i^2-\left(\sum_{i=1}^nx_i\right)^2/n\right]\bar{y}}},  
\end{eqnarray}
where $\bar{y}=\sum_{i=1}^n\frac{y_i}{n}$. 

\textit{It is worthwhile noticing that the formulas for the logistic regression (\ref{logistic}) and for the Poisson regression (\ref{poisson}) are very similar to the Pearson correlation coefficient (\ref{r})}. This is a cornerstone feature of the score test for these two regression models that will reduce the computational burden significantly.

\item With strictly positive response values, Gamma regression is an ordinarily selected model, whose log-likelihood is given by 
$$\ell_1=\sum_{i=1}^n\left[\alpha -\frac{y_i}{\mu_i}-\log{\left(\mu_i\right)}+\alpha\log{\left(\alpha y_i\right)}-\log{\left(\alpha\right)}\right],$$ where $\mu_i=e^{a+bx_i}$. The score test for Gamma regression has the following formula
\begin{eqnarray} \label{gama}
S_{Ga} = \frac{\sum_{i=1}^nx_i-\frac{\sum_{i=1}^ny_ix_i}{\hat{\alpha}/\hat{\beta}}}{\sqrt{\sum_{i=1}^nx_i^2/\hat{\alpha}}}, 
\end{eqnarray}
where $\hat{\alpha}$ and $\hat{\beta}$ are the MLE estimates of the Gamma regression under $H_0$. 

\item With count data that exhibit overdispersion (variance is greater than the mean), the negative binomial regression is more suitable than the Poisson regression that assumes the dispersion parameter is 1 (mean is equal to the variance). The relevant log-likelihood is given by $$\ell_1=\sum_{i=1}^n\left[\log{\Gamma\left(y_i+r\right)}-\log{\left(y_i!\right)}-\log{\left(r\right)}+r\log{\left(\frac{r}{r+\mu_i}\right)}+y_i\log{\left(\frac{\mu_i}{r+\mu_i}\right)} \right],$$ where $\mu_i=e^{a+bx_i}$. The corresponding score test is given by
\begin{eqnarray} \label{nb}
S_{NB} = \frac{\hat{p}\sum_{i=1}^nx_iy_i-\left(1-\hat{p}\right) \hat{r}\sum_{i=1}^nx_i}{\sqrt{\hat{p}^2\left(\bar{y}+\bar{y}^2/\hat{r}\right)\sum_{i=1}^nx_i^2}},  
\end{eqnarray}
where $\bar{y}$ is the sample mean, $\hat{p}$ and $\hat{r}$ are the MLE estimates of the Negative Binomial regression under $H_0$. 

\item Beta regression is appropriate for responses that lie within $\left(0, 1\right)$ with the log-likelihood being 
\begin{eqnarray*}
\ell_1 = & & \sum_{i=1}^n\left\lbrace\log{\Gamma\left(\phi\right)}-\log{\Gamma\left(\mu_i\phi\right)}-\log{\Gamma\left[\left(1-\mu_i\right)\phi\right]}+\left(\mu_i\phi-1\right)\log{\left(y_i\right)} +\right.  
\\ & & +\left. \left[\left(1-\mu_i\right)\phi-1\right]\log{\left(1-y_i\right)}  \right\rbrace,
\end{eqnarray*}
where $\mu_i=\frac{1}{1+e^{-a-bx_i}}$. The relevant score test is given by
\begin{eqnarray} \label{beta}
S_{Be} = \frac{\sum_{i=1}^nx_i\log{\frac{y_i}{1-y_i}}-\sum_{i=1}^nx_i\left[\psi(\hat{\alpha})-\psi(\hat{\beta})\right]}{\sqrt{\sum_{i=1}^nx_i^2\left[\psi'(\hat{\alpha})+\psi'(\hat{\beta})\right]}},
\end{eqnarray}
where $\hat{\alpha}$ and $\hat{\beta}$ are the MLE estimates of the Beta regression under $H_0$, $\psi(.)$ and $\psi'(.)$ are the digamma and trigamma functions respectively. 

\item An alternative to Gamma regression is the Weibull regression, that is mainly used in biostatistics. Its log-likelihood is given by $$\ell_1=\sum_{i=1}^n\left[\log{\left(\kappa\right)}-\log{\left(\lambda_i\right)}+\left(\kappa-1\right)\log{\left(\frac{y_i}{\lambda_i}\right)}- \left(\frac{y_i}{\lambda_i}\right)^{\kappa}\right],$$ where $\lambda_i=e^{a+bx_i}$. The relevant score test takes the following form
\begin{eqnarray} \label{weib}
S_{Weib} = \frac{\frac{\sum_{i=1}^nx_iy_i^{\hat{\kappa}}}{\hat{\lambda}^{\hat{\kappa}}} - \sum_{i=1}^nx_i}{\sqrt{\sum_{i=1}^nx_i^2}},
\end{eqnarray}
where $\hat{\kappa}$ and $\hat{\lambda}$ are the MLE estimates of the Weibull regression under $H_0$. 

\end{itemize}

\subsection{Welch's $t$-test for binary responses}
When the response is binary, the Welch's $t$-test \cite{welch1951} can also be used and it's test statistic is given by
\begin{eqnarray} \label{tw}
T_w=\frac{\bar{x}_1-\bar{x}_2 }{\sqrt{\frac{s_1^2}{n_1}+\frac{s_2^2}{n_2}}},
\end{eqnarray} 
where $\bar{x}_1$ and $\bar{x}_2$ denote the two sample means and $s_1^2$ and $s_2^2$ are the two sample variances. Under $H_0$, $T_w \sim t_{\nu}$, with $t_{\nu}$ denoting the $t$ distribution with $\nu$ degrees of freedom and $\nu$ is given by \citep{satterthwaite1946, welch1951}
\begin{eqnarray}  \label{nu}
\nu \simeq \frac{\left(\frac{s_1^2}{n_1} + \frac{s_2^2}{n_2}\right)^2}{\frac{s_1^4}{n_1^2(n_1-1)} + \frac{s_2^4}{n_2^2(n_2-1)}}.
\end{eqnarray}

According to \citep{boulesteix2007} this is one of the standard approaches for such cases. To the best of the authors' knowledge this test is not frequently employed by variable selection algorithms and has gone unnoticed. One possible reason could be that no one has performed simulation studies or empirical evaluation studies and show its its undermined value. The non parametric alternative, Wilcoxon-Mann-Whitney test is not suggested because it tends to inflate the type I error \citep{tsagris2018b}. 

\section{Related work}
The issue of computational efficiency has drawn the research interest of many researchers. \citet{sikorska2013} proposed an efficient approximation test for logistic regression, which can be used to obtain thousands of p-values, but it is not as computationally efficient as the score test.  \citet{redden2004} proposed a fast method, based on logistic regression, for obtaining the p-values of many median regressions. Obtaining the p-value of a logistic regression is much faster than obtaining the p-value of a median regression. When large sample sizes are available, adoption of the score test can make their method computationally extremely efficient compared to conducting numerous logistic regressions. 

Computer nowadays have made parallel computations easier and more efficient. \cite{tsamardinos2019} took advantage of the parallel computing and adopted the Forward Backward with Early Dropping algorithm \citep{borboudakis2019} for big (and massive) data. Parallel computing takes place not only across the predictor variables, but across the observations as well. The observations are split into folds and a logistic regression model is fitted in each fold. The results are then meta-analytically combined. This process produces accurate results with hundreds of thousands of observations and can lead to substantial improvements in terms of execution time, up to 10 times faster. The computational reduction during the univariate filtering though is not comparable to the one achieved by the score test. 

On a different direction, \citet{erdogdu2019} proved that, asymptotically, the beta coefficients of generalised linear models are proportional to the beta coefficients of a linear model. Our simulation studies provided evidence that this holds true for other regression models also, e.g. Weibull regression. Despite fitting a linear model is much cheaper than fitting a logistic regression model for instance, the computational savings are not as significant as one would think. Finding the proportionality factor, requires application of the Newton-Raphson or the golden-ratio algorithm that go through the whole dataset at each step. Undoubtedly, this process is faster than simply fitting many (non-linear) regression models, yet, it is not as efficient as performing many score tests.  

Another direction is to use sub-samples of the data instead of the whole dataset \citep{park2018} with the trade-off of this strategy being accuracy. According to \citet{park2018}, their proposed method, that uses a portion of the data, can speed-up the maximum likelihood estimation of the model from 6 up to 629 times compared to using the full dataset while guaranteeing the same model predictions, with $95\%$ probability. The score test on the contrary, will be shown to return nearly the same results as the log-likelihood ratio test at a level of more than $99\%$ similarity, with large sample sizes.

\section{Monte Carlo simulations} \label{simulations}
Three regression models will be examined, logistic regression, Gamma regression and Beta regression. Since the score test for the logistic regression is very similar to the Pearson correlation, the latter will be excluded from this regression. In all cases, the four axes of comparison or four metrics are:
a) \textit{Computational cost}, b) \textit{Type I error}, c) \textit{Correlation of the p-values} and d) \textit{Agreement in the decision (reject/not reject $H_0$)}.

\subsection{Example 1: Logistic regression}
Binary response values were generated from a Bernoulli distribution with various probabilities of success $p=(0.1, 0.2, 0.3, 0.4, 05)$ while $d=500$ random predictor variables were generated from a standard normal distribution. The sample size varied from $10,000$ up to $1,000,000$. For each combination of probability of success and sample size the aforementioned four metrics were computed. This process was repeated $10$ times and the average performance metrics are reported.

\begin{figure}[!h]
\includegraphics[scale=0.72, trim = 0 20 0 0]{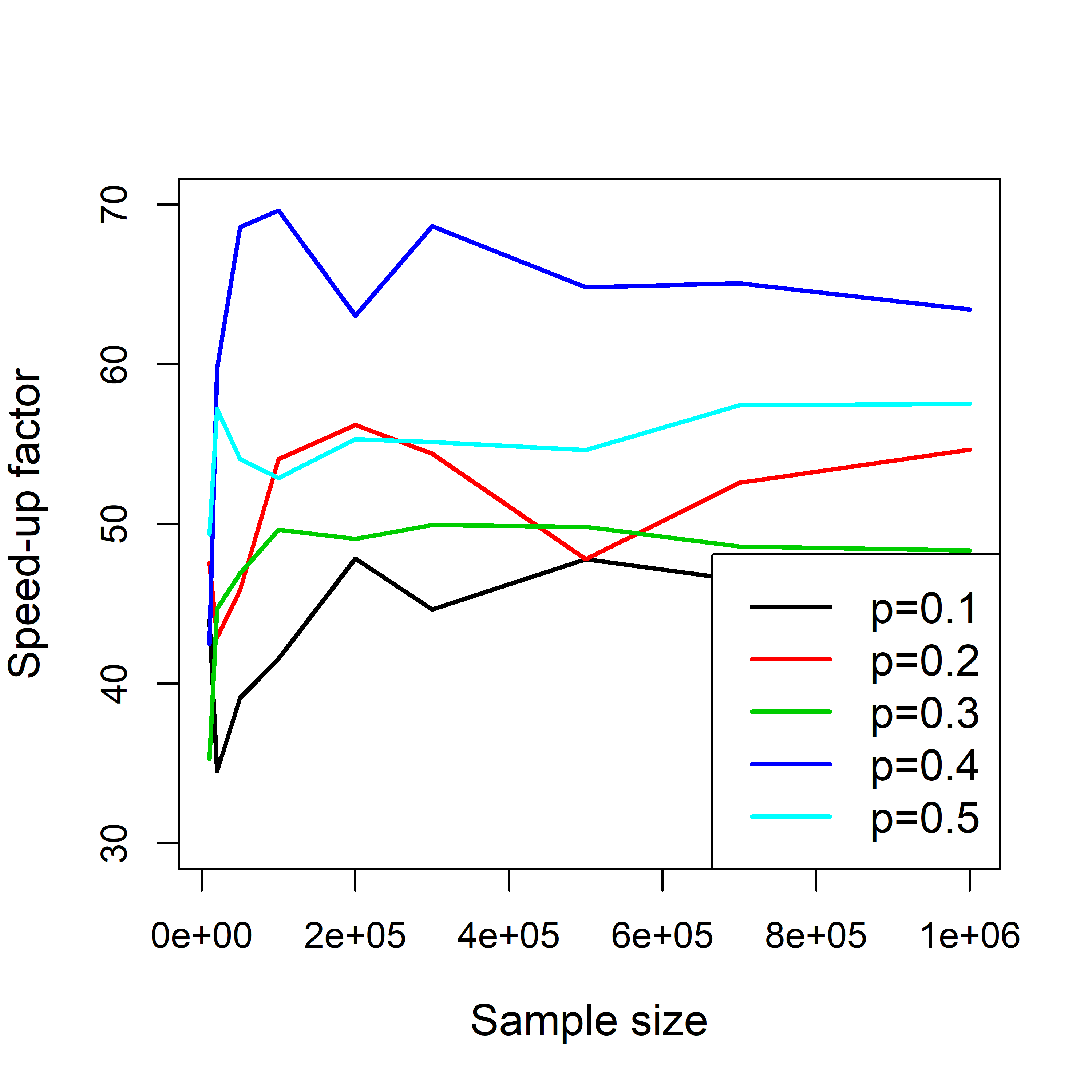}
\centering
\caption{\textbf{Logistic regression: Speed-up factor of the $\Lambda$ test against the $S^2$ test}. This is an estimate of how many times the $\Lambda$ test is slower than the $S^2$ test. \label{logi_time2}}
\end{figure}

Table \ref{logi_time} shows computational cost (in seconds) of each test for the $500$ predictor variables for different sample sizes. The computational cost of both tests increases with the sample size, with the log-likelihood ratio test requiring up to $6$ minutes with large sample sizes, while the score test never exceeds $6$ seconds. Figure \ref{logi_time2} presents the speed-up factor\footnote{The number of times the log-likelihood ratio test is slower than the score test.} across the various probabilities of success as a function of the sample size. The log-likelihood ratio test is between $30$ to $70$ times slower than the score test.

Table \ref{logi_alpha} contains the estimated type I error for both tests. These are in close agreement and when the sample size is $20,000$ or higher the estimated errors have the same value up to the 3rd digit. The correlation of the p-values of the two tests is perfect when the sample size is $20,000$ or larger (see Table \ref{logi_others}). The percentage of agreement in the decision of rejection of the $H_0$ is also perfect (see Table \ref{logi_others}) for the same sample sizes. 

\begin{scriptsize}
\begin{table}[!ht]
\caption{\textbf{Logistic regression}: \textbf{Computational cost} (in seconds) of the log-likelihood ratio test ($\Lambda$) and the score test ($S^2$) for different sample sizes and probabilities of success. The fastest method is highlighted in bold. \label{logi_time}}
\begin{center}
\begin{tabular}{c|cc|cc|cc|cc|cc} \toprule
      & \multicolumn{10}{c}{Probability of success}    \\   \hline 
      &  \multicolumn{2}{c}{$p=0.1$}  &  \multicolumn{2}{c}{$p=0.2$}   &  \multicolumn{2}{c}{$p=0.3$}   &  \multicolumn{2}{c}{$p=0.4$}   &  \multicolumn{2}{c}{$p=0.5$}    \\ \hline  
Sample size   & $\Lambda$  & $S^2$ & $\Lambda$  & $S^2$ & $\Lambda$  & $S^2$ & $\Lambda$  & $S^2$
              & $\Lambda$  & $S^2$ \\  \midrule
1x$10^4$  &  1.72  &  \textbf{0.04}  &  1.52  &  \textbf{0.03}  &  1.3  &  \textbf{0.04}  &  1.83  &  \textbf{0.04}  &  1.78  &  \textbf{0.04}  \\
2x$10^4$  &  4.11  &  \textbf{0.12}  &  3.09  &  \textbf{0.07}  &  3.08  &  \textbf{0.07}  &  3.88  &  \textbf{0.06}  &  3.15  &  \textbf{0.05}  \\
5x$10^4$  &  13.5  &  \textbf{0.34}  &  9.26  &  \textbf{0.2}  &  7.79  &  \textbf{0.17}  &  11.18  &  \textbf{0.16}  &  8.38  &  \textbf{0.15}  \\
1x$10^5$  &  25.99  &  \textbf{0.63}  &  18.27  &  \textbf{0.34}  &  14.99  &  \textbf{0.3}  &  20.2  &  \textbf{0.29}  &  16.55  &  \textbf{0.31}  \\
2x$10^5$  &  58.86  &  \textbf{1.23}  &  38.16  &  \textbf{0.68}  &  32.34  &  \textbf{0.66}  &  43.31  &  \textbf{0.69}  &  33.85  &  \textbf{0.61}  \\
3x$10^5$  &  87.67  &  \textbf{1.96}  &  58.00  &  \textbf{1.07}  &  48.85  &  \textbf{0.98}  &  62.26  &  \textbf{0.91}  &  50.67  &  \textbf{0.92}  \\
5x$10^5$  &  107.04  &  \textbf{2.24}  &  104.1  &  \textbf{2.18}  &  81.51  &  \textbf{1.64}  &  105.2  &  \textbf{1.62}  &  86.32  &  \textbf{1.58}  \\
7x$10^5$  &  183.1  &  \textbf{3.94}  &  132.19  &  \textbf{2.51}  &  113.28  &  \textbf{2.33}  &  146.48  &  \textbf{2.25}  &  123.37  &  \textbf{2.15}  \\
1x$10^6$  &  254.99  &  \textbf{5.62}  &  178.27  &  \textbf{3.26}  &  156.82  &  \textbf{3.24}  &  207.02  &  \textbf{3.26}  &  178.18  &  \textbf{3.10}  \\ \bottomrule
\end{tabular}
\end{center}
\end{table}
\end{scriptsize} 

The Welch's $t$-test produces similar results to the score test and hence are not presented. The speed-up factors ranged from $34$ up to $59$ and the estimated type I errors were almost identical. The correlation of the log-likelihood ratio test p-values with the Welch's $t$-test p-values was always $1$ and the percentage of agreement in rejecting the null hypothesis was either $0.998$, $0.999$ or $1$.

\begin{scriptsize}
\begin{table}
\caption{\textbf{Logistic regression: Estimated type I error} of the log-likelihood ratio test ($\Lambda$) and the score test ($S^2$) for different sample sizes and probabilities of success.  \label{logi_alpha}}
\begin{center}
\begin{tabular}{c|cc|cc|cc|cc|cc} \toprule
        & \multicolumn{10}{c}{Probability of success}    \\  \hline   
Sample  &  \multicolumn{2}{c}{$p=0.1$}  &  \multicolumn{2}{c}{$p=0.2$}   &  \multicolumn{2}{c}{$p=0.3$}   &  \multicolumn{2}{c}{$p=0.4$}   &  \multicolumn{2}{c}{$p=0.5$}   \\ \hline  
size      & $\Lambda$  & $S^2$ & $\Lambda$  & $S^2$ & $\Lambda$  & $S^2$ & $\Lambda$  & $S^2$
          & $\Lambda$  & $S^2$ \\  \midrule
1x$10^4$  &  0.056  &  0.056  &  0.053  &  0.053  &  0.053  &  0.052  &  0.052  &  0.051  &  0.049  &  0.049 \\
2x$10^4$  &  0.051  &  0.051  &  0.055  &  0.055  &  0.049  &  0.049  &  0.048  &  0.048  &  0.048  &  0.048 \\
5x$10^4$  &  0.053  &  0.053  &  0.050  &  0.050  &  0.052  &  0.052  &  0.053  &  0.053  &  0.045  &  0.045  \\
1x$10^5$  &  0.051  &  0.051  &  0.052  &  0.052  &  0.050  &  0.050  &  0.047  &  0.047  &  0.048  &  0.048  \\
2x$10^5$  &  0.048  &  0.048  &  0.053  &  0.053  &  0.052  &  0.052  &  0.050  &  0.050  &  0.050  &  0.050  \\
3x$10^5$  &  0.047  &  0.047  &  0.048  &  0.048  &  0.051  &  0.051  &  0.049  &  0.049  &  0.057  &  0.057  \\
5x$10^5$  &  0.052  &  0.052  &  0.052  &  0.052  &  0.049  &  0.049  &  0.048  &  0.048  &  0.046  &  0.046  \\
7x$10^5$  &  0.050  &  0.050  &  0.048  &  0.048  &  0.054  &  0.054  &  0.052  &  0.052  &  0.047  &  0.047  \\
1x$10^6$  &  0.050  &  0.050  &  0.052  &  0.052  &  0.045  &  0.045  &  0.048  &  0.048  &  0.051  &  0.051  \\  \bottomrule
\end{tabular}
\end{center}
\end{table}
\end{scriptsize} 

\begin{scriptsize}
\begin{table}[!ht]
\caption{\textbf{Logistic regression}: \textbf{Correlation of the $\Lambda$ and $S^2$ test p-values and percentage of agreement in rejecting $\bf H_0$} for different sample sizes and probabilities of success. \label{logi_others}}
\begin{center}
\begin{tabular}{c|ccccc|c|ccccc} \toprule
         &  \multicolumn{5}{c}{\textbf{Correlation of the p-values}}  &  & \multicolumn{5}{c}{\textbf{Percentage of agreement}} \\ \hline
  & \multicolumn{5}{c}{Probability of success ($p$)} &  & \multicolumn{5}{c}{Probability of success ($p$)}  \\ \midrule
Sample size      &  0.1  &  0.2  &  0.3  &  0.4  &  0.5  &  &  0.1  &  0.2  &  0.3  &  0.4  &  0.5  \\ \hline \hline
1x$10^4$  &  1  &  1  &  1  &  1  &  1  &  &  1  &  1  &  0.999  &  0.999  &  1  \\  
$\geq$ 2x$10^4$  &  1  &  1  &  1  &  1  &  1  &  &  1  &  1  &  1  &  1  &  1  \\  \bottomrule
\end{tabular}
\end{center}
\end{table}
\end{scriptsize} 

\subsection{Example 2: Gamma regression}
\begin{figure}[!h]
\begin{tabular}{cc}
\includegraphics[scale=0.6, trim = 0 20 0 0]{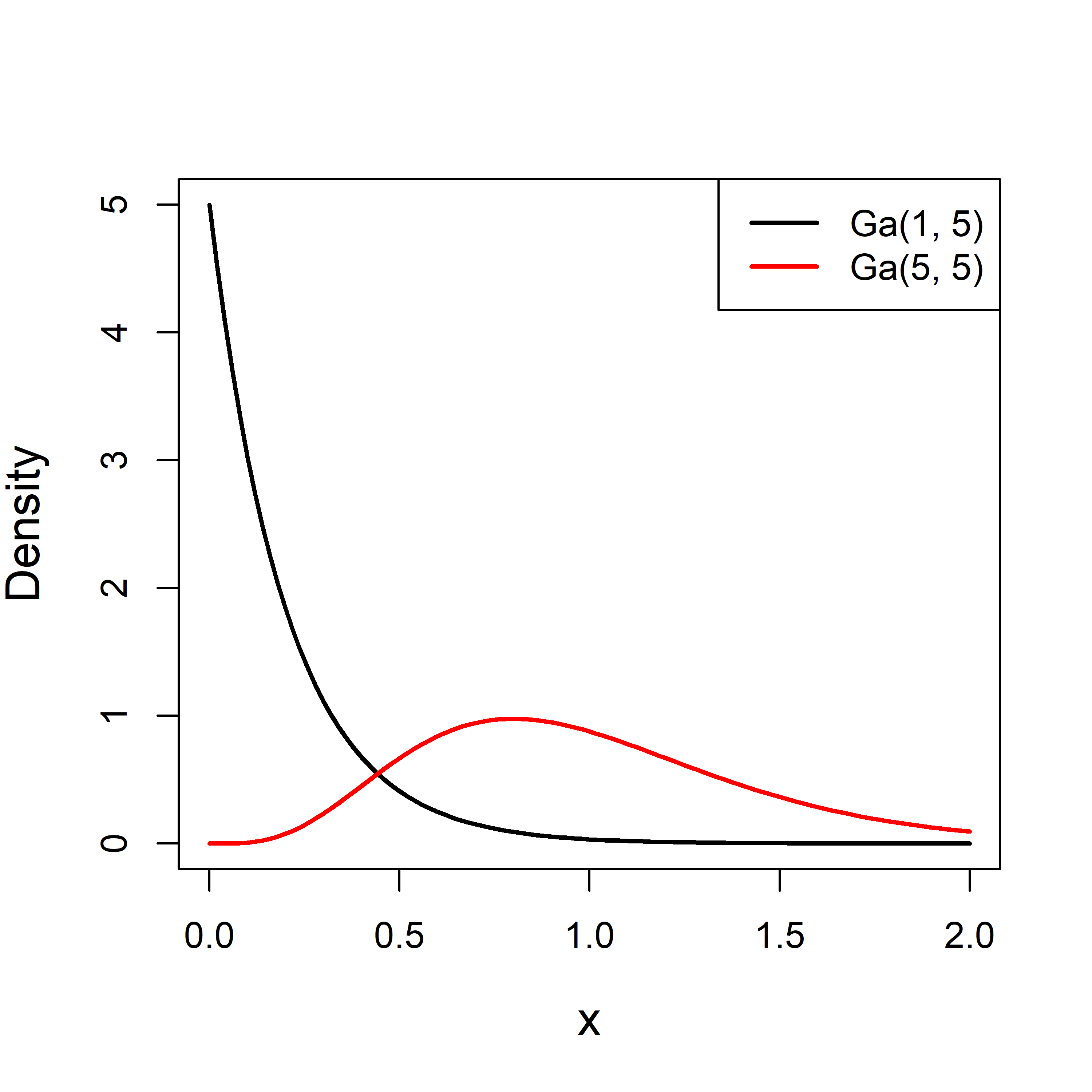} &
\includegraphics[scale=0.6, trim = 0 20 20 0]{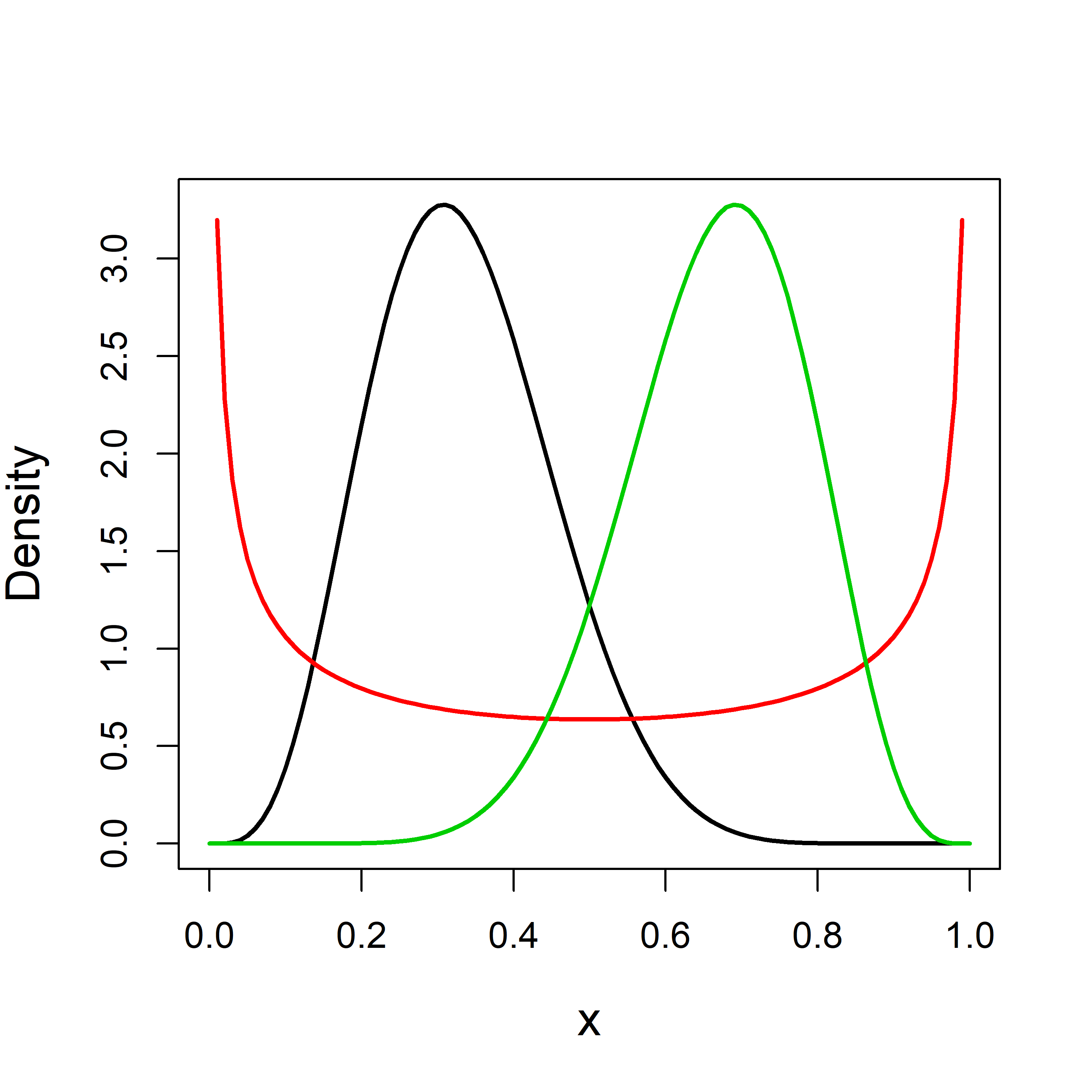} \\
\textbf{(a)  Gamma distribution}   &  \textbf{(b) Beta distribution} 
\end{tabular}
\caption{(a) \textbf{Gamma densities} for different values of $\alpha$ and $\beta$. 
(b) \textbf{Beta densities} for different values of $\alpha$ and $\beta$. \textbf{\textemdash}: Be(5, 10), \textcolor{red}{\textbf{\textemdash}:} Be(0.5, 0.5), 
\textcolor{green}{\textbf{\textemdash}:} Be(10, 5). \label{densities} }
\end{figure}

Response values were generated from a Ga($1$, $5$) and a Ga($5$, $5$) (the shape of these densities appear in Figure \ref{densities}(a)), and for each of them $d=500$ random predictor variables were generated from standard normal distribution. The sample sizes varied again from $10,000$ up to $1,000,000$ and for each Gamma distribution and sample size the four performance metrics were computed and averaged over $10$ repetitions. The results are presented in Tables \ref{gamma_time}, \ref{gamma_alpha_1} and \ref{gamma_others}. 

Table \ref{gamma_time}) summarizes the computational cost of the log-likelihood ratio test and of the score test. The computational cost of the log-likelihood ratio test for large sample sizes is as high as 6 minutes, whereas for the score test it never exceeded the 4 seconds. The speed-up factor varies from $52$ up to $93$. For both Gamma distributions considered the computational benefit (speed-up factor) of the score test is large and then decreases until it reaches a plateau at about 55, for sample sizes equal to hundreds of thousands. 

The estimated type I errors of both tests are in close agreement as can be seen in Table \ref{gamma_alpha_1}. There is a very close agreement between the log-likelihood ratio test and the score test even for sample sizes equal to $10,000$. Their estimated type I errors become equal when the sample sizes are $50,000$ or more for both Gamma distributions. 

The correlation of the p-values (see Table \ref{gamma_others}) reaches 1 for sample sizes equal to or greater than $100,000$. The percentage of agreement in rejecting the $H_0$ or not (see Table \ref{gamma_others}) also reaches 1 for the sample sizes equal to or greater than $100,000$. Nonetheless, the correlation is satisfactorily high for smaller sample sizes and never drops below $0.999$.

\begin{scriptsize}
\begin{table}[!ht]
\caption{\textbf{Gamma regression}: \textbf{Computational cost} (in seconds) of the log-likelihood ratio test ($\Lambda$) and the score test ($S^2$) for different sample sizes and parameter values with $d=500$ predictor variables. Fastest method is highlighted in bold. The speed-up factor columns depict the number of times $\Lambda$ is slower than $S^2$. \label{gamma_time}}
\begin{center}
\begin{tabular}{c|cc|c|cc|c} \toprule
        & \multicolumn{6}{c}{Gamma parameters}    \\   \hline 
  &  \multicolumn{3}{c}{$\alpha=1$, $\beta=5$}  &  \multicolumn{3}{c}{$\alpha=5$, $\beta=5$}   \\ \hline  
Sample size     & $\Lambda$  & $S^2$ & Speed-up factor & $\Lambda$  & $S^2$  & Speed-up factor \\  \midrule
1x$10^4$  & 1.43     &  \textbf{0.02}  & 71.50  &  1.87    &  \textbf{0.02}  & 93.50 \\
2x$10^4$  &  2.83    &  \textbf{0.04}  & 70.75  &  3.61    &  \textbf{0.04}  & 90.25 \\
5x$10^4$  &  7.24    &  \textbf{0.10}  & 72.40  &  8.68    &  \textbf{0.11}  & 78.91 \\
1x$10^5$  &  16.21   &  \textbf{0.23}  & 70.48  &  18.85   &  \textbf{0.30}  & 62.83 \\
2x$10^5$  &  28.74   &  \textbf{0.56}  & 51.32  &  34.28   &  \textbf{0.60}  & 57.13 \\
3x$10^5$  &  47.08   &  \textbf{0.87}  & 54.11  &  53.21   &  \textbf{0.96}  & 55.43 \\
5x$10^5$  &  84.17   &  \textbf{1.47}  & 57.23  &  82.26   &  \textbf{1.58}  & 52.06 \\
7x$10^5$  &  132.08  &  \textbf{2.43}  & 54.36  &  103.29  &  \textbf{1.97}  & 52.43 \\
1x$10^6$  &  188.98  &  \textbf{3.40}  & 55.58  &  143.54  &  \textbf{2.54}  & 56.51 \\ \bottomrule
\end{tabular}
\end{center}
\end{table}
\end{scriptsize}  

\begin{scriptsize}
\begin{table}[!ht]
\caption{\textbf{Gamma regression: Estimated type I error} of the log-likelihood ratio test ($\Lambda$) and the score test ($S^2$) for different sample sizes and parameter values with $d=100$ predictor variables. \label{gamma_alpha_1}}
\begin{center}
\begin{tabular}{c|cc|cc} \toprule
     & \multicolumn{4}{c}{Gamma parameters}    \\   \hline 
     &  \multicolumn{2}{c}{$\alpha=1$, $\beta=5$}  &  \multicolumn{2}{c}{$\alpha=5$, $\beta=5$}  \\ \hline  
Sample size         & $\Lambda$  & $S^2$ & $\Lambda$  & $S^2$ \\  \midrule
1x$10^4$ & 0.048  &  0.049  &  0.046  &  0.045 \\
2x$10^4$ & 0.048  &  0.048  &  0.050  &  0.050 \\
5x$10^4$ & 0.052  &  0.052  &  0.049  &  0.049 \\
1x$10^5$ & 0.056  &  0.055  &  0.055  &  0.055 \\
2x$10^5$ & 0.048  &  0.047  &  0.047  &  0.047 \\
3x$10^5$ & 0.049  &  0.050  &  0.050  &  0.050 \\
5x$10^5$ & 0.046  &  0.046  &  0.052  &  0.052 \\
7x$10^5$ & 0.055  &  0.055  &  0.046  &  0.046 \\
1x$10^6$ & 0.050  &  0.050  &  0.049  &  0.049 \\  \bottomrule
\end{tabular}
\end{center}
\end{table}
\end{scriptsize} 

\begin{scriptsize}
\begin{table}[!ht]
\caption{\textbf{Gamma regression}: \textbf{Correlation of the $\Lambda$ and $S^2$ test p-values and percentage of agreement in rejecting $\bf H_0$} for different sample sizes and probabilities of success. \label{gamma_others}}
\begin{center}
\begin{tabular}{c|cc|cc} \toprule
      &  \multicolumn{2}{c}{\textbf{Correlation of p-values}}  &  \multicolumn{2}{c}{\textbf{Percentage of agreement}}   \\ \hline 
      & \multicolumn{4}{c}{Gamma parameters}  \\ \hline
Sample size  & $\alpha=1$, $\beta=5$    & $\alpha=5$, $\beta=5$ & $\alpha=1$, $\beta=5$  & $\alpha=1$, $\beta=5$ \\  \midrule
1x$10^4$ & 0.999  &  0.999  &  0.999  &  0.999 \\
2x$10^4$ & 0.999  &  1  &  0.999  &  0.999 \\
5x$10^4$ & 0.999  &  1  &  0.999  &  1 \\
$\geq$ 1x$10^5$   &  1  &  1  &  1  &  1 \\ \bottomrule
\end{tabular}
\end{center}
\end{table}
\end{scriptsize} 

\subsection{Example 3: Beta regression}
The response values this time were generated from a Be($\alpha$, $\beta$) and Figure \ref{densities}(b) shows the probability density function of the Beta distribution with the three different pairs of parameters used. In this scenario a) $p=100$ and b) $p=500$ random predictor variables were generated from standard normal distribution, while the sample sizes varied from $100$ up to $20,000$. The reason was that the score test was shown to be size correct even for small sample sizes \citep{cribari2014}. Beta regression, implemented in the R package \textbf{betareg} \citep{zeileis2010}, is not implemented in C++ but utilizes the R built-in function \textit{optim} and hence the computational cost increases considerably with sample size. 

The average duration (in seconds) of $100$ univariate Beta regressions and of $100$ score tests appears in Table \ref{beta_time_1}. The speed-up factors are more than $4,000$, meaning that $100$ Beta regressions can be more than $4,000$ times slower than $100$ score tests. The estimated type I errors (see Table \ref{beta_alpha_1}) are nearly the same. Note that this time the sample size was only as large as $20,000$, as the time required for the Beta regressions increases with the sample size. A similar picture is taken by examining the computational cost in Table \ref{beta_time_2} and the estimated type I error in Table \ref{beta_alpha_2} for the case of $500$ predictor variables. The computational cost is dramatically smaller than performing $500$ Beta regressions in R. The speed-up factor ranges from $154$ up to $5809$, indicating that performing many Beta regressions can be thousands of times slower than performing many score test. 

Surprisingly enough, the score test is faster than the Pearson correlation coefficients (see Table \ref{beta_time_2}). However, Table \ref{beta_alpha_2} shows that the estimated type I error of the score test and of the Pearson correlation coefficient do not fully agree even for sample sizes equal to $1,000,000$. 

In order to see whether this disagreement was significant and to see what are the possible implications, the probability of identifying the most significant variable was computed for the score test and for the Pearson correlation coefficient. The motivation behind is because the (generalised) Orthogonal Matching Pursuit \citep{tsagris2018a} algorithm selects the most significant variable in the first step. If the score test and the Pearson correlation coefficient agree in the most significant variable, then their type I error differences can be deemed negligible. In this case, one predictor variable ($x_i$) was randomly chosen from the $500$ predictor variables. The response values were then generated from $Be\left(\frac{1}{1+e^{-x_i}}, 1\right)$. The results, presented in Table \ref{beta_prob}, show that when the sample sizes exceed $500$ there is perfect agreement, in detecting the most statistically significant variable, between the score test and the Pearson correlation coefficient.

\begin{scriptsize}
\begin{table}[!ht]
\caption{\textbf{Beta regression}: \textbf{Computational cost} (in seconds) of the log-likelihood ratio test ($\Lambda$) and the score test ($S^2$) for different sample sizes and parameter values with $p=100$ predictor variables. Fastest method is highlighted in bold. The speed-up factor columns depict the number of times $\Lambda$ is slower than $S^2$. \label{beta_time_1}}
\begin{center}
\begin{tabular}{c|cc|c|cc|c|cc|c} \toprule
      & \multicolumn{9}{c}{Beta parameters}    \\   \hline 
    &  \multicolumn{3}{c}{$\alpha=5$, $\beta=10$}  &  \multicolumn{3}{c}{$\alpha=0.5$, $\beta=0.5$}   &  \multicolumn{3}{c}{$\alpha=10$, $\beta=5$}   \\ \midrule  
\multirow{2}{*}{Sample size} &  &  & Speed-up &  &  & Speed-up &  &  & Speed-up \\
        &  $\Lambda$  &  $S^2$  &  factor   &   $\Lambda$  &  $S^2$  & factor  & $\Lambda$  & $S^2$ & factor \\ \midrule
100    & 1.54  & \textbf{0.01} & 154  & 1.64  & \textbf{0.01} & 164  & 1.73  & \textbf{0.01} & 173 \\
500    & 2.73  & \textbf{0.01} & 273  & 2.63  & \textbf{0.01} & 263  & 3.19  & \textbf{0.01} & 319 \\
1,000  & 4.37  & \textbf{0.01} & 437  & 4.11  & \textbf{0.01} & 411  & 5.71  & \textbf{0.01} & 571 \\
5,000  & 21.18 & \textbf{0.01} & 2118 & 16.61 & \textbf{0.01} & 1661 & 24.11 & \textbf{0.01} & 2411 \\
10,000 & 54.99 & \textbf{0.01} & 5499 & 35.46 & \textbf{0.01} & 1773 & 58.09 & \textbf{0.01} & 5809 \\
20,000 & 88.68 & \textbf{0.02} & 4434 & 66.03 & \textbf{0.02} & 3315 & 91.62 & \textbf{0.02} & 4581 \\  \bottomrule
\end{tabular}
\end{center}
\end{table}
\end{scriptsize} 

\begin{scriptsize}
\begin{table}[!ht]
\caption{\textbf{Beta regression: Estimated type I error} of the log-likelihood ratio test ($\Lambda$) and the score test ($S^2$) for different sample sizes and parameter values with $p=100$ predictor variables. \label{beta_alpha_1}}
\begin{center}
\begin{tabular}{c|cc|cc|cc} \toprule
     & \multicolumn{6}{c}{Beta parameters}    \\   \hline 
     &  \multicolumn{2}{c}{$\alpha=5$, $\beta=10$}  &  \multicolumn{2}{c}{$\alpha=0.5$, $\beta=0.5$} &  \multicolumn{2}{c}{$\alpha=10$, $\beta=5$}   \\ \midrule 
Sample size   & $\Lambda$  & $S^2$ & $\Lambda$  & $S^2$ & $\Lambda$  & $S^2$ \\  \hline \hline
100     &  0.052  &  0.048  &  0.064  &  0.061  &  0.054  &  0.052  \\
500     &  0.051  &  0.048  &  0.054  &  0.053  &  0.037  &  0.036  \\
1,000   &  0.050  &  0.051  &  0.048  &  0.047  &  0.059  &  0.057  \\
5,000   &  0.067  &  0.066  &  0.044  &  0.044  &  0.046  &  0.046  \\
10,000  &  0.047  &  0.047  &  0.041  &  0.041  &  0.048  &  0.048  \\
20,000  &  0.052  &  0.052  &  0.054  &  0.054  &  0.047  &  0.047  \\  \bottomrule
\end{tabular}
\end{center}
\end{table}
\end{scriptsize}

\begin{scriptsize}
\begin{table}[!ht]
\caption{\textbf{Beta regression}: \textbf{Computational cost} (in seconds) of the score test ($S^2$) and the Pearson correlation coefficient test ($Z$) for different sample sizes and parameter values with $d=500$ predictor variables. Fastest method is highlighted in bold. \label{beta_time_2}}
\begin{center}
\begin{tabular}{c|cc|cc|cc} \toprule
   & \multicolumn{6}{c}{Beta parameters}    \\   \hline 
   &  \multicolumn{2}{c}{$\alpha=5$, $\beta=10$}  &  \multicolumn{2}{c}{$\alpha=0.5$, $\beta=0.5$}   &  \multicolumn{2}{c}{$\alpha=10$, $\beta=5$}   \\ \hline
Sample size   & $S^2$  & $Z$ & $S^2$  & $Z$ & $S^2$  & $Z$ \\  \midrule
100       &  0.00  &  0.00  &  0.00  &  0.00  &  0.00  &  0.00  \\
500       &  0.00  &  0.00  &  0.00  &  0.00  &  0.00  &  0.00  \\
1,000     &  0.00  &  0.00  &  0.00  &  0.01  &  0.00  &  0.00  \\
5,000     &  0.02  &  0.02  &  \textbf{0.01}  &  0.02  &  \textbf{0.01}  &  0.02  \\
1x$10^4$  &  \textbf{0.03}  &  0.04  &  \textbf{0.02}  &  0.03  &  \textbf{0.02}  &  0.03  \\
2x$10^4$  &  \textbf{0.05}  &  0.06  &  \textbf{0.05}  &  0.06  &  \textbf{0.05}  &  0.06  \\
5x$10^4$  &  \textbf{0.12}  &  0.17  &  \textbf{0.12}  &  0.15  &  \textbf{0.12}  &  0.16  \\
1x$10^5$  &  \textbf{0.29}  &  0.41  &  \textbf{0.23}  &  0.29  &  \textbf{0.25}  &  0.30  \\
2x$10^5$  &  \textbf{0.56}  &  0.68  &  \textbf{0.47}  &  0.60  &  \textbf{0.47}  &  0.61  \\
5x$10^5$  &  \textbf{1.36}  &  1.74  &  \textbf{1.19}  &  1.53  &  \textbf{1.34}  &  1.66  \\
7x$10^5$  &  \textbf{1.94}  &  2.54  &  \textbf{1.65}  &  2.13  &  \textbf{2.02}  &  2.41  \\
1x$10^6$  &  \textbf{2.48}  &  3.09  &  \textbf{2.72}  &  3.17  &  \textbf{2.88}  &  3.45  \\  \bottomrule
\end{tabular}
\end{center}
\end{table}
\end{scriptsize} 

\begin{scriptsize}
\begin{table}[!ht]
\caption{\textbf{Beta regression: Estimated type I error} of the score test ($S^2$) and the Pearson correlation coefficient test ($Z$) for different sample sizes and parameter values with $d=500$ predictor variables. \label{beta_alpha_2}}
\begin{center}
\begin{tabular}{c|cc|cc|cc} \toprule
     & \multicolumn{6}{c}{Beta parameters}    \\   \hline 
     &  \multicolumn{2}{c}{$\alpha=5$, $\beta=10$}  &  \multicolumn{2}{c}{$\alpha=0.5$, $\beta=0.5$}   &  \multicolumn{2}{c}{$\alpha=10$, $\beta=5$}   \\ \midrule  
Sample size      & $S^2$   & $Z$ & $S^2$  & $Z$ & $S^2$  & $Z$ \\  \hline \hline
100       &  0.048  &  0.046  &  0.052  &  0.049  &  0.052  &  0.048  \\
500       &  0.055  &  0.058  &  0.045  &  0.047  &  0.047  &  0.047  \\
1,000     &  0.055  &  0.054  &  0.054  &  0.051  &  0.055  &  0.055  \\
5,000     &  0.049  &  0.048  &  0.055  &  0.054  &  0.051  &  0.052  \\
1x$10^4$  &  0.052  &  0.053  &  0.053  &  0.051  &  0.052  &  0.051  \\
2x$10^4$  &  0.051  &  0.052  &  0.048  &  0.050  &  0.051  &  0.051  \\
5x$10^4$  &  0.051  &  0.055  &  0.045  &  0.045  &  0.051  &  0.053  \\
1x$10^5$  &  0.046  &  0.047  &  0.045  &  0.047  &  0.055  &  0.052  \\
2x$10^5$  &  0.052  &  0.049  &  0.049  &  0.047  &  0.045  &  0.046  \\
5x$10^5$  &  0.054  &  0.054  &  0.044  &  0.046  &  0.049  &  0.050  \\
7x$10^5$  &  0.055  &  0.054  &  0.054  &  0.053  &  0.047  &  0.047  \\
1x$10^6$  &  0.045  &  0.046  &  0.048  &  0.051  &  0.045  &  0.045  \\  \bottomrule
\end{tabular}
\end{center}
\end{table}
\end{scriptsize} 

\begin{scriptsize}
\begin{table}[!ht]
\caption{\textbf{Beta regression: Estimated probability of identifying the most significant predictor variable} of the score test ($S^2$) and the Pearson correlation coefficient test ($Z$) for different sample sizes and parameter values with $p=500$ predictor variables. The highest probability is highlighted in bold. \label{beta_prob}}
\begin{center}
\begin{tabular}{c|cc|cc|cc} \toprule
      & \multicolumn{6}{c}{Beta parameters}    \\   \hline 
      &  \multicolumn{2}{c}{$\alpha=5$, $\beta=10$}  &  \multicolumn{2}{c}{$\alpha=0.5$, $\beta=0.5$}   &  \multicolumn{2}{c}{$\alpha=10$, $\beta=5$}   \\ \hline  
Sample size & $S^2$   & $Z$ & $S^2$  & $Z$ & $S^2$  & $Z$ \\  \midrule
100         &  \textbf{0.92}  &  0.64  &  \textbf{0.90}  &  0.62  &  \textbf{0.88}  &  0.50  \\
$\geq 500$  &  1.00  &  1.00  &  1.00  &  1.00  &  1.00  &  1.00  \\  \bottomrule
\end{tabular}
\end{center}
\end{table}
\end{scriptsize} 

\section{Examples with real data} \label{real_examples}
The computational cost of the log-likelihood and of the score test, the correlation of their corresponding p-values and the percentage of agreement in rejecting/not rejecting the $H_0$ were next assessed using real data. Monte Carlo studies are based on simulating the predictor variables and the response variable from parametric models followed by parametric regression models. Hence, the data generating mechanism is expected to be recovered with large sample sizes. On the contrary, examples with real data will illustrate the robustness of the aforementioned tests to model miss-specification, since real data are very unlikely to obey any parametric model assumptions. 

Two datasets were downloaded from the \href{https://archive.ics.uci.edu/ml/index.php}{UC Irvine Machine Learning Repository}, namely the \textit{Gisette} dataset and the \textit{Online News Popularity} dataset. Both datasets have a binary response and are thus suitable for logistic regression. The first dataset is a handwritten digit recognition problem where the goal is to separate the highly confusible digits "4" and "9". This dataset is one of five datasets of the NIPS 2003 feature selection challenge \citep{guyon2005} and contains $5,999$ binary observations and $5,000$ predictor variables. The second dataset summarizes a heterogeneous set of features about articles published by Mashable in a period of two years \citep{fernandes2015} with the goal of predicting the popularity in social networks. The popularity of online news is often measured by considering the number of interactions in the Web and social networks (e.g., number of shares, likes and comments). The authors have binarised the popularity using a threshold of $1,400$ shares and thus have turned the regression problem into a classification problem. This dataset contains $39,644$ observations and $64$ predictor variables.

\begin{itemize}
\item \textbf{Logistic regression and score test}. In order to obtain a better and more accurate picture of the computational cost, the execution time and the relevant performance metrics were measured 10 times. Each time a bootstrap sample was generated containing the response vector and the predictor variables matrix (the pairing was not distorted).  

\item \textbf{Gamma regression and score test}. Since the response values are binary, non negative continuous random values were generated from a mixture of a Weibull and a folded normal distribution with the mixing proportion being equal to $50\%$. This process was repeated 10 times. 

\item \textbf{Beta regression and score test}. Similarly to Gamma regression, percentages were generated from a mixture of a logistic normal distribution and a simplex distribution with the mixing proportion being equal to $50\%$. This process was repeated 5 times only for the first dataset (and 10 times for the second dataset), because fitting thousands of Beta regressions is computationally highly expensive.
\end{itemize}

The performance metrics that were computed are a) the computational cost of the log-likelihood ratio test and of the score test, b) the correlation of their corresponding p-values and c) the percentage of agreement in rejecting/not rejecting the $H_0$. The average numbers of all metrics are reported in Table \ref{real}, corroborating the evidence of the simulations for the case of logistic regression. The first dataset (Gisette) contains $5999$ observations and this explains why the correlation between the log-likelihood ratio p-values and score test p-values is $0.999$. The second dataset (Online) contains 39644 observations and this is why the correlation of the p-values is $1$. The same conclusions were drawn for Welch's $t$-test. The results agree with the simulation studies also, for the Gamma and Beta regressions. The correlation of the p-values is only $0.997$ even for the second dataset (Online). Table \ref{gamma_others} reported that the correlation of the p-values of the score test and the log-likelihood ratio tests requires tens of thousands of observations. Finally the computational advantage of the score test (and of the Welch's $t$-test) over the log-likelihood ratio test is again evident for all three types of regressions.

\begin{scriptsize}
\begin{table}[!ht]
\caption{\textbf{Real data examples}: \textbf{Computational cost (in seconds) of $\Lambda$ and $S^2$ tests, speed-up factor (No of times $\Lambda$ is slower than $S^2$) correlation of their p-values and percentage of agreement in rejecting $\bf H_0$} for the two datasets and the three regression models. \label{real}}
\begin{center}
\begin{tabular}{c|c|ccc|cc|cc|cc} \toprule
&     &  \multicolumn{3}{c}{\textbf{Computational}} & \multicolumn{2}{c}{\textbf{Speed-up}}  & \multicolumn{2}{c}{\textbf{Correlation}} &  \multicolumn{2}{c}{\textbf{Percent of}}  \\ 
&     &  \multicolumn{3}{c}{\textbf{cost}} & \multicolumn{2}{c}{\textbf{factor}}  & \multicolumn{2}{c}{\textbf{of p-values}} &  \multicolumn{2}{c}{\textbf{agreement}}  \\  \hline
Regression   &  Dataset    &  $\Lambda$ & $S^2$ & Welch & $S^2$ & Welch & $S^2$ & Welch & $S^2$ & Welch  \\  \midrule 
\multirow{2}{*}{Logistic}  & Gisette   & 19.875    &  0.239 & 0.242  & 83.16    & 82.13  & 0.999   & 0.999  & 0.991 & 0.991 \\
                           & Online    & 1.438     &  0.020 & 0.016  & 71.90    & 89.88  & 1       & 0.999  & 1     & 1     \\  \hline
\multirow{2}{*}{Gamma}     & Gisette   & 13.913    &  0.251 &        & 55.43    &        & 0.995   &        & 0.973 &      \\
                           & Online    & 0.908     &  0.008 &        & 113.50   &        & 0.997   &        & 0.983 &      \\  \hline
\multirow{2}{*}{Beta}      & Gisette   & 1208.310  &  0.250 &        & 4833.24  &        & 1       &        & 0.998 &      \\
                           & Onine     & 70.988    & 0.112  &        & 6172.87  &        & 1       &        & 1     &       \\  \bottomrule\end{tabular}
\end{center}
\end{table}
\end{scriptsize} 

\begin{figure}[htp]
\centering
{\begin{tabular}{cc}
\textbf{Gisette dataset}   &   \textbf{Online News Popularity dataset}  \\
\includegraphics[scale = 0.55, trim = 40 0 0 0]{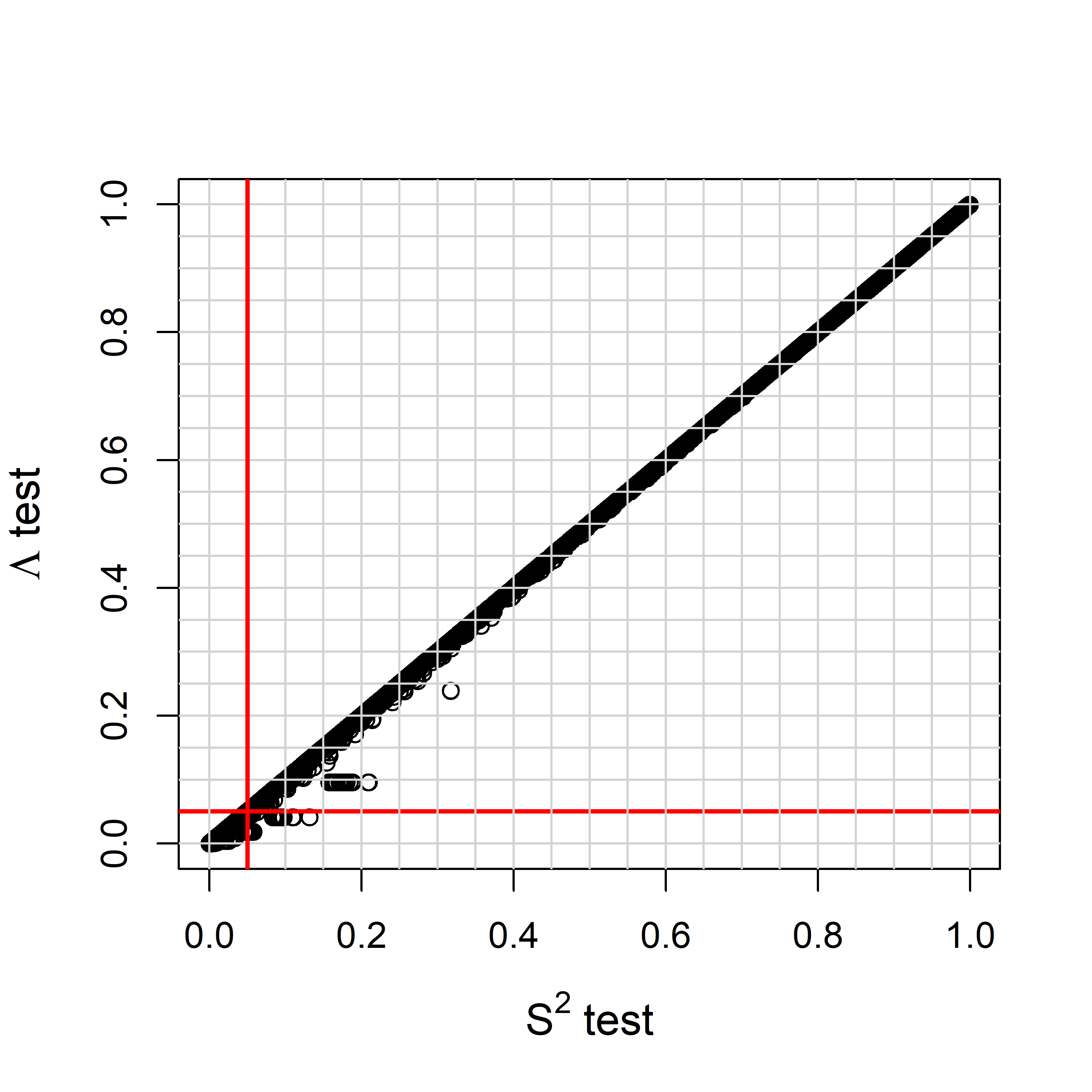} &
\includegraphics[scale = 0.55, trim = 40 0 0 0]{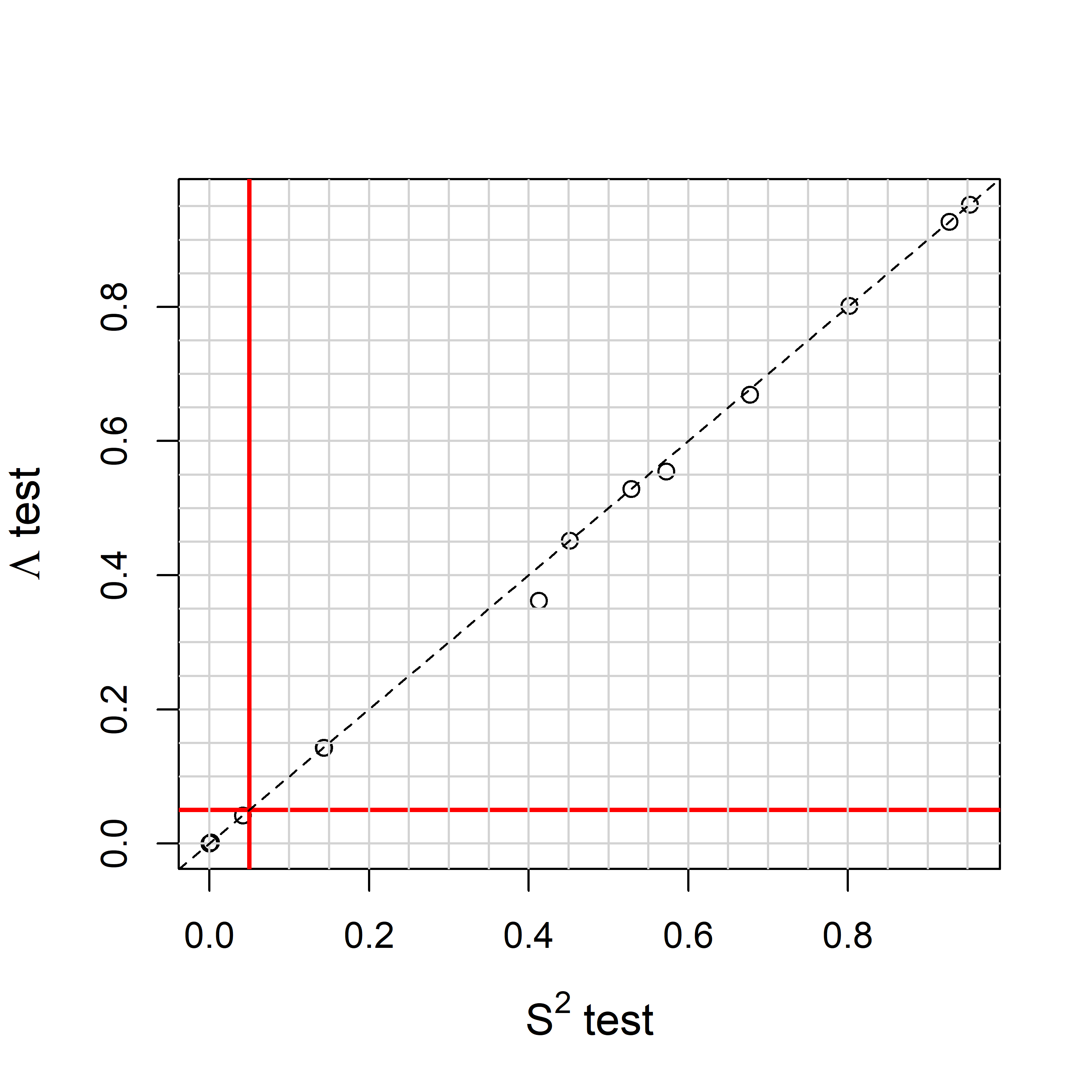} \\
\textbf{(a) $S^2$ p-values versus $\Lambda$ p-values}  & 
\textbf{(b) $S^2$ p-values versus $\Lambda$ p-values}  \\
\includegraphics[scale = 0.55, trim = 40 0 0 0]{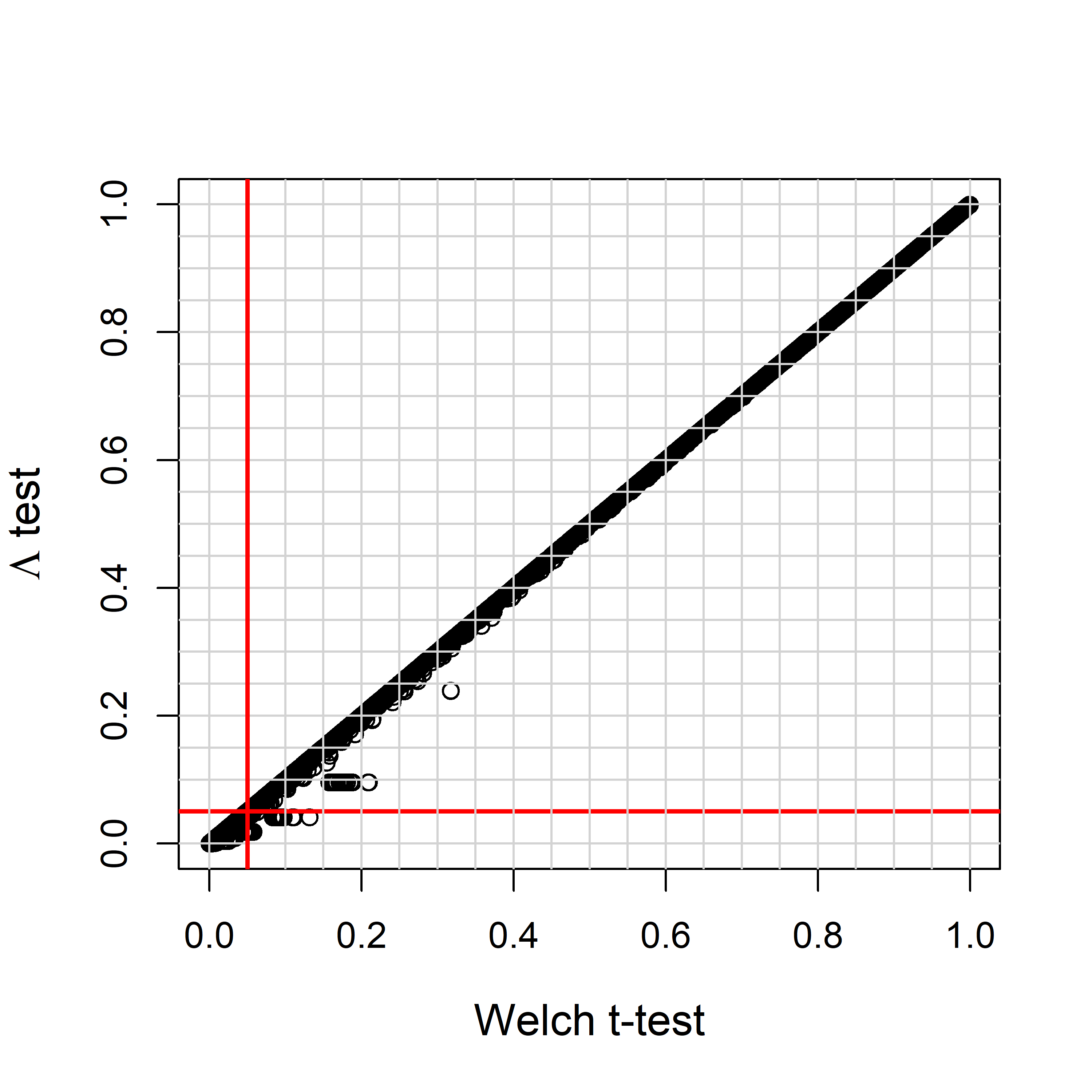} &
\includegraphics[scale = 0.55, trim = 40 0 0 0]{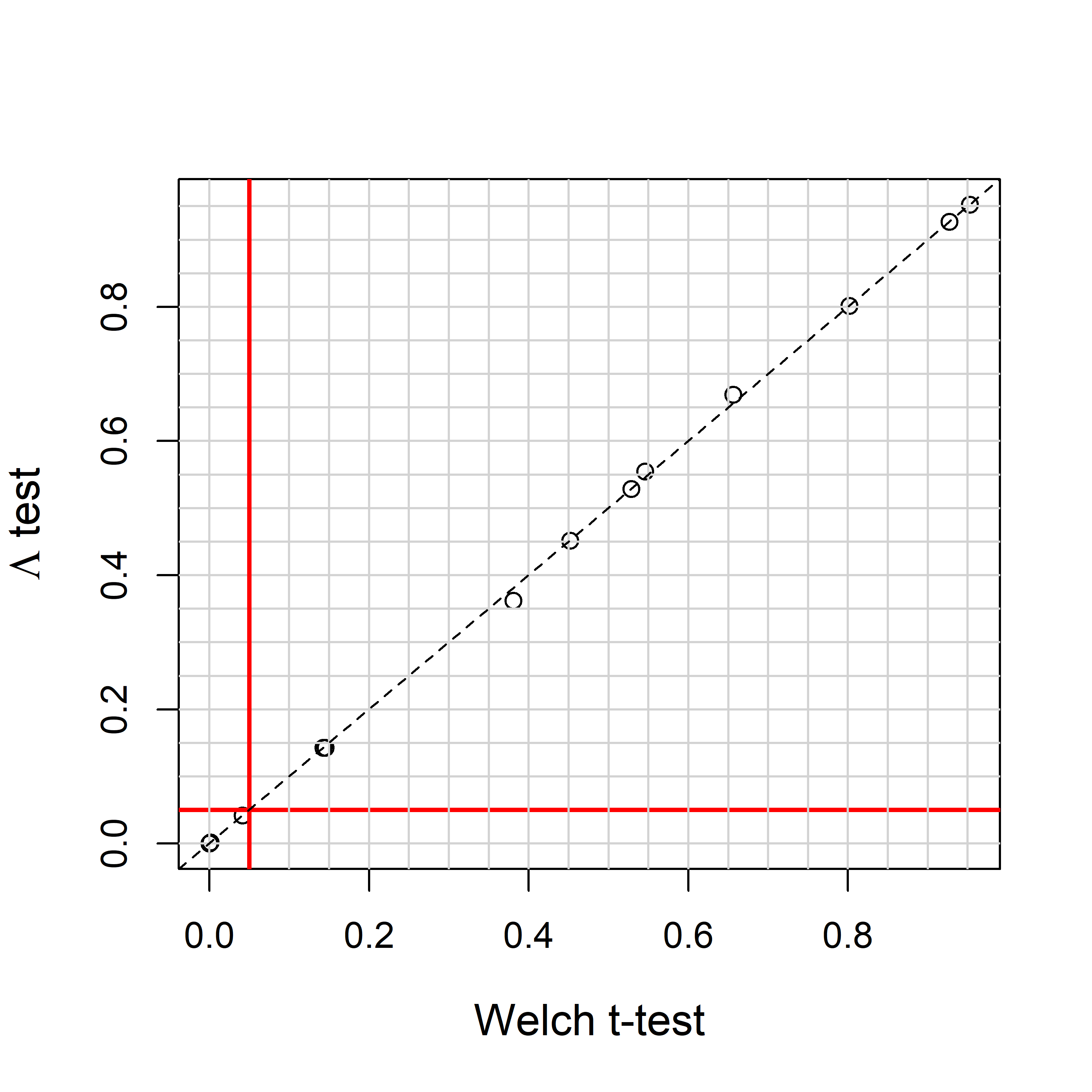} \\
\textbf{(c) Welch's $t$-test p-values versus $\Lambda$ p-values}  & 
\textbf{(d) Welch's $t$-test p-values versus $\Lambda$ p-values}
\end{tabular}}
\caption{\textbf{Scatter plot of the score test ($S^2$) p-values and of the Welch's $t$-test p-values versus the log-likelihood ratio test ($\Lambda$) p-values} (using logistic regression). The dashed line refers to the $45^{\circ}$ line that passes through the origin. The red lines delimit the rejection region at the $5\%$ significance level for each test.}
\label{pvalues} 
\end{figure}

For the case of sample sizes being less than a few tens of thousands, a heuristic was tested. Assume that the significance threshold is set to $5\%$. The score test is first performed and the variables whose p-value is less than $10\%$ are stored and the log-likelihood ratio test is applied to these variables only. This heuristic was applied to more than $50$ gene expression datasets, but the results did not support this strategy hence are not presented here. The computational savings were significant, but on the other hand many variables identified as significant by the log-likelihood ratio test were not identified as such by the score test.

\section{Conclusions} \label{conclusions}
The score test was suggested as a faster alternative to log-likelihood ratio test that involves fitting many simple (with one predictor) regression models. Score test's only requirement, in order to be equivalent to the log-likelihood ratio test, is large sample size. This might sound like a disadvantage at first, but is actually an advantage. With large scale or massive data, computational cost becomes a serious problem and score test solves this problem effectively. 

The score test and the Pearson correlation coefficient when used for univariate filtering were shown to be computationally extremely efficient when compared to the log-likelihood ratio test and produced exactly the same results with large sample sizes ( $n>10,000$) for logistic regression and Gamma regression. In addition, the Welch's $t$-test produced almost identical results to the score test. Hence, with large sample sized data or massive and big data, the score test could substitue the log-likelihood ratio test, and for logistic regression the Welch's $t$-test is another option.

For Beta distributed response values, the Pearson correlation coefficient and the score test did not reach $100\%$ agreement for smaller sample sizes. The interesting conclusion though is that the score test is size correct even for small sample sizes corroborating the findings of \citet{cribari2014}. This implies that the score test could replace the log-likelihood ratio test even for small sample sizes with Beta distributed response values.

Another conclusion this paper has reached to, is that despite R being rather "slow" (in comparison to Python or Matlab), with the proper computations it becomes extremely fast. The general advice "\textit{\textbf{It's your algorithm}}" suits the results of this paper. Continuing with this, we would like to inform the reader that many score and log-likelihood ratio tests have been implemented in the R packages \textbf{Rfast}\citep{rfast2019} and \textbf{Rfast2} \citep{rfast22019}. Furthermore, we are working towards improving the computational efficiency of the score test.

Due to the paper's space limitations not many regression cases could be covered. For instance, Poisson and negative binomial and Weibull regression for which the formulas of the score test were provided. The case of multinomial regression was not examined either, for which Welch's $F$-test for multiple samples \citep{welch1951} can be an alternative to the log-likelihood ratio test, with computational cost nearly equal to that of the score test and results of similar accuracy. 

Future research includes assessement of the the score test in general, not only for univariate filtering purposes. The Forward Backward with Early Dropping \citep{borboudakis2019} variable selection algorithm performs numerous log-likelihood ratio tests. Addressing the computational cost associated with big data, \citep{tsamardinos2019} proposed a meta-analytic formulation of those tests. Adaptation of the score test could result to higher computational savings because fewer regression models will be built.

\bibliographystyle{apalike}

\end{document}